# Mathematical Modeling of Soliton's Evolution in Generalized Quantum Hydrodynamics

Boris V. Alexeev
Moscow Academy of Fine Chemical Technology (MITHT)
Prospekt Vernadskogo, 86, Moscow 119570, Russia
B.Alexeev@ru.net

Abstract

This paper addresses the fundamental principles of generalized Boltzmann physical kinetics, as a part of non-local physics. It is shown that the theory of transport processes (including quantum mechanics) can be considered in the frame of unified theory based on the non-local physical description. The paper can be considered also as comments and prolongation of the materials published in the known author`s monograph (Boris V. Alexeev, Generalized Boltzmann Physical Kinetics. Elsevier. 2004). The theory leads to solitons as typical formations in the generalized quantum hydrodynamics.

## 1. About basic principles of the Generalized Boltzmann Physical Kinetics and non-local physics.

As it is shown (see, for example [1, 2]) the theory of transport processes (including quantum mechanics) can be considered in the frame of unified theory based on the non-local physical description. In particular the generalized hydrodynamic equations represent an effective tool for solving problems in the very vast area of physical problems. The following results in this paper can be considered as prolongation of the theory published by author in [1, 2]. Nevertheless for simplicity, we will consider fundamental methodic aspects from the qualitative standpoint and avoid excessively cumbersome formulas. Let us discuss the correlation between the generalized Boltzmann physical kinetics and the laws of conservation. A rigorous description is found, for example, in the monograph [1].

Transport processes in open dissipative systems are considered in physical kinetics. Therefore, the kinetic description is inevitably related to the system diagnostics. Such an element of diagnostics in the case of theoretical description in physical kinetics is the concept of the physically infinitely small volume (**PhSV**). The correlation between theoretical description and system diagnostics is well-known in physics. Suffice it to recall the part played by test charge in electrostatics or by test circuit in the physics of magnetic phenomena. The traditional definition of **PhSV** contains the statement to the effect that the **PhSV** contains a sufficient number of particles for introducing a statistical description; however, at the same time, the **PhSV** is much smaller than the volume $V$ of the physical system under consideration; in a first approximation, this leads to local approach in investigating the transport processes. It is assumed in classical hydrodynamics that local thermodynamic equilibrium is first established within the **PhSV**, and only after that the transition occurs to global thermodynamic equilibrium if it is at all possible for the system under study. Let us consider the hydrodynamic description in more detail from this point of view. Assume that we have two neighboring physically infinitely small volumes **PhSV$_1$** and **PhSV$_2$** in a nonequilibrium system. The one-particle distribution function (DF) $f_{sm,1}(\mathbf{r}_1, \mathbf{v}, t)$ corresponds to the volume **PhSV$_1$**, and the function $f_{sm,2}(\mathbf{r}_2, \mathbf{v}, t)$ — to the volume **PhSV$_2$**. It is assumed in a first approximation that $f_{sm,1}(\mathbf{r}_1, \mathbf{v}, t)$ does not vary within **PhSV$_1$**, same as $f_{sm,2}(\mathbf{r}_2, \mathbf{v}, t)$ does not vary within the neighboring volume **PhSV$_2$**. It is this assumption of locality that is implicitly contained in the Boltzmann equation (BE). However, the assumption is too crude. Indeed, a particle on the boundary between two volumes, which experienced the last collision in **PhSV$_1$** and moves toward **PhSV$_2$**, introduces information about the $f_{sm,1}(\mathbf{r}_1, \mathbf{v}, t)$ into the neighboring volume **PhSV$_2$**. Similarly, a particle on the boundary between two volumes, which experienced the last collision in **PhSV$_2$** and moves toward **PhSV$_1$**, introduces information about the DF $f_{sm,2}(\mathbf{r}_2, \mathbf{v}, t)$ into the neighboring volume **PhSV$_1$**. The relaxation over translational degrees of freedom of particles of like masses occurs during several collisions. As a result, "Knudsen layers" are formed on the boundary between neighboring physically infinitely small volumes, the characteristic dimension of which is of the order of path length. Therefore, a correction must be introduced into the DF in the **PhSV**, which is proportional to the mean time between collisions and to the substantive derivative of the DF being measured (rigorous derivation is given in [1]). Let a particle of finite radius be characterized as before by the position **r** at the instant of time *t* of its center of mass moving at velocity **v**. Then, the situation is possible where, at some instant of time *t*, the particle is located on the interface between two volumes. In so doing, the lead effect is possible (say, for **PhSV$_2$**), when the center of mass of particle moving to the neighboring volume **PhSV$_2$** is still in **PhSV$_1$**. However, the delay effect takes place as well, when the center of mass of particle moving to the neighboring volume (say, **PhSV$_2$**) is already located in **PhSV$_2$** but a part of the particle still belongs to **PhSV$_1$**. This entire complex of effects defines non-local effects in space and time.

The physically infinitely small volume (**PhSV**) is an *open* thermodynamic system *for any division of macroscopic system by a set of PhSVs.* However, the BE [3, 4]

$$Df/Dt = J^B, \qquad (1.1)$$

where $J^B$ is the Boltzmann collision integral and $D/Dt$ is a substantive derivative, fully ignores non-local effects and contains only the local collision integral $J^B$. The foregoing nonlocal effects are insignificant only in equilibrium systems, where the kinetic approach changes to methods of statistical mechanics.

This is what the difficulties of classical Boltzmann physical kinetics arise from. Also a weak point of the classical Boltzmann kinetic theory is the treatment of the dynamic properties of interacting particles. On the one hand, as follows from the so-called "physical" derivation of the BE, Boltzmann particles are regarded as material points; on the other hand, the collision integral in the BE leads to the emergence of collision cross sections.

The rigorous approach to derivation of kinetic equation relative to one-particle DF $f$ ($KE_f$) is based on employing the hierarchy of Bogoliubov equations. Generally speaking, the structure of $KE_f$ is as follows:

$$\frac{Df}{Dt} = J^B + J^{nl}, \tag{1.2}$$

where $J^{nl}$ is the non-local integral term. An approximation for the second collision integral is suggested by me in *generalized* Boltzmann physical kinetics,

$$J^{nl} = \frac{D}{Dt}\left(\tau \frac{Df_1}{Dt}\right). \tag{1.3}$$

Here, $\tau$ is the mean time *between* collisions of particles, which is related in a hydrodynamic approximation with dynamical viscosity $\mu$ and pressure $p$,

$$\tau\, p = \Pi \mu, \tag{1.4}$$

where the factor $\Pi$ is defined by the model of collision of particles: for neutral hard-sphere gas, $\Pi = 0.8$ [5]. All of the known methods of deriving kinetic equation relative to one-particle DF lead to approximation (1.3), including the method of many scales, the method of correlation functions, and the iteration method. One can draw an analogy with the Bhatnagar–Gross–Krook (BGK) approximation for local integral $J^B$,

$$J^B = \frac{f_1^{(0)} - f_1}{\tau}, \tag{1.5}$$

the popularity of which in the case of Boltzmann collision integral is explained by the colossal simplification attained when using this approximation. The order of magnitude of the ratio between the second and first terms of the right-hand part of Eq. (1.2) is $Kn^2$, at high values of Knudsen number, these terms come to be of the same order. It would seem that, at low values of Knudsen number corresponding to hydrodynamic description, the contribution by the second term of the right-hand part of Eq. (1.2) could be ignored. However, this is not the case. Upon transition to hydrodynamic approximation (following the multiplication of the kinetic equation by invariants collision and subsequent integration with respect to velocities), the Boltzmann integral part goes to zero, and the second term of the right-hand part of Eq. (1.2) *does not go to zero* after this integration and produces a contribution of the same order in the case of generalized Navier–Stokes description. From the mathematical standpoint, disregarding the term containing a small parameter with higher derivative is impermissible. From the physical standpoint, the arising additional terms proportional to viscosity correspond to Kolmogorov small-scale turbulence; the fluctuations are tabulated [1, 6]. It turns out that the integral term $J^{nl}$ is important from the standpoint of the theory of transport processes at both low and high values of Knudsen number. Note the treatment of GBE from the standpoint of fluctuation theory,

$$Df^a/Dt = J^B, \tag{1.6}$$

$$f^a = f - \tau\, Df/Dt \tag{1.7}$$

Equations (1.6) and (1.7) have a correct free-molecule limit. Therefore, $\tau\, Df/Dt$ is a fluctuation of distribution function, and the notation (1.6) disregarding (1.7) renders the BE open. From the standpoint of fluctuation theory, Boltzmann employed the simplest closing procedure

$$f^a = f. \tag{1.8}$$

Fluctuation effects occur in any open thermodynamic system bounded by a control surface transparent to particles. GBE (1.6) leads to generalized hydrodynamic equations [1], for example, to the continuity equation

$$\frac{\partial \rho^a}{\partial t} + \frac{\partial}{\partial \mathbf{r}} \cdot (\rho \mathbf{v}_0)^a = 0, \tag{1.9}$$

where $\rho^a$, $\mathbf{v}_0{}^a$, $(\rho\mathbf{v}_0)^a$ are calculated in view of non-locality effect in terms of gas density $\rho$, hydrodynamic velocity of flow $\mathbf{v}_0$, and density of momentum flux $\mathbf{v}_0$; for locally Maxwellian distribution, $\rho^a$, $(\rho\mathbf{v}_0)^a$ are defined by the relations

$$(\rho - \rho^a)/\tau = \frac{\partial \rho}{\partial t} + \frac{\partial}{\partial \mathbf{r}} \cdot (\rho\mathbf{v}_0), \quad (\rho\mathbf{v}_0 - (\rho\mathbf{v}_0)^a)/\tau = \frac{\partial}{\partial t}(\rho\mathbf{v}_0) + \frac{\partial}{\partial \mathbf{r}} \cdot \rho\mathbf{v}_0\mathbf{v}_0 + \vec{I} \cdot \frac{\partial p}{\partial \mathbf{r}} - \rho\mathbf{a}, \quad (1.10)$$

where $\vec{I}$ is a unit tensor, and $\mathbf{a}$ is the acceleration due to the effect of mass forces. In the general case, the parameter $\tau$ is the non-locality parameter; in quantum hydrodynamics, its magnitude is defined by the "time-energy" uncertainty relation [2]. The violation of Bell's inequalities [7, 8] is found for local statistical theories, and the transition to non-local description is inevitable.

## 2. Generalized hydrodynamic equations (GHE) and quantum hydrodynamics.

The following conclusion of principal significance can be done from the previous consideration [2]:
1. Madelung's quantum hydrodynamics is equivalent to the Schroedinger equation (SE) and leads to description of the quantum particle evolution in the form of Euler equation and continuity equation.
2. SE is consequence of the Liouville equation as result of the local approximation of non-local equations.
3. Generalized Boltzmann physical kinetics leads to the strict approximation of non-local effects in space and time and after transmission to the local approximation leads to parameter $\tau$, which on the quantum level corresponds to the uncertainty principle "time-energy".
4. GHE lead to SE as a deep particular case of the generalized Boltzmann physical kinetics and therefore of non-local hydrodynamics.

In the following we intend to obtain the soliton's type of solution of the generalized hydrodynamic equations. Much more examples of the theory applications can be found in [9]. On the first step of this investigation we use assumptions which were introduced in implicit form in quantum mechanics of Madelung and Schroedinger. On the finalized step I shall use the non-stationary 1D model with taking into account the energy equation, external forces and non-locality parameter $\tau$ defined by the "time-energy" uncertainty relation of Heisenberg.

Non-stationary 1D GHE will be solved [1, 2]:
(Continuity equation)

$$\frac{\partial}{\partial t}\left\{\rho - \tau^{(0)}\left[\frac{\partial \rho}{\partial t} + \frac{\partial}{\partial x}(\rho v_0)\right]\right\} + \frac{\partial}{\partial x}\left\{\rho v_0 - \tau^{(0)}\left[\frac{\partial}{\partial t}(\rho v_0) + \frac{\partial}{\partial x}(\rho v_0^2) + \frac{\partial p}{\partial x} + \rho\frac{\partial U}{\partial x}\right]\right\} = 0, \quad (2.1)$$

(Momentum equation)

$$\frac{\partial}{\partial t}\left\{\rho v_0 - \tau^{(0)}\left[\frac{\partial}{\partial t}(\rho v_0) + \frac{\partial}{\partial x}(\rho v_0^2) + \frac{\partial p}{\partial x} + \rho\frac{\partial U}{\partial x}\right]\right\} +$$
$$+ \frac{\partial U}{\partial x}\left[\rho - \tau^{(0)}\left(\frac{\partial \rho}{\partial t} + \frac{\partial}{\partial x}(\rho v_0)\right)\right] + \quad (2.2)$$
$$+ \frac{\partial}{\partial x}\left\{\rho v_0^2 + p - \tau^{(0)}\left[\frac{\partial}{\partial t}(\rho v_0^2 + p) + \frac{\partial}{\partial x}(\rho v_0^3 + 3pv_0) + 2\rho v_0\frac{\partial U}{\partial x}\right]\right\} = 0,$$

(Energy equation)

$$\frac{\partial}{\partial t}\left\{\rho v_0^2 + 3p - \tau^{(0)}\left[\frac{\partial}{\partial t}\left(\rho v_0^2 + 3p\right) + \frac{\partial}{\partial x}\left(\rho v_0^3 + 5pv_0\right) + 2\rho v_0 \frac{\partial U}{\partial x}\right]\right\} +$$

$$+ \frac{\partial}{\partial x}\left\{\rho v_0^3 + 5pv_0 - \tau^{(0)}\left[\frac{\partial}{\partial t}\left(\rho v_0^3 + 5pv_0\right) + \frac{\partial}{\partial x}\left(\rho v_0^4 + 8pv_0^2 + 5\frac{p^2}{\rho}\right) + \frac{\partial U}{\partial x}\left(3\rho v_0^2 + 5p\right)\right]\right\} +$$

$$+ 2\frac{\partial U}{\partial x}\left\{\rho v_0 - \tau^{(0)}\left[\frac{\partial}{\partial t}(\rho v_0) + \frac{\partial}{\partial x}\left(\rho v_0^2 + p\right) + \rho\frac{\partial U}{\partial x}\right]\right\} = 0, \qquad (2.3)$$

where $U$ is potential energy of unit mass, then the force $F = -\frac{\partial U}{\partial x}$, $F$ is the force acting on the mass unit. On the mentioned above the first step of investigation we follow the Schroedinger – Madelung assumptions. Namely,

a) The influence of time delay effects is neglected.
b) The influence of terms containing the pressure $p$ in the hydrodynamic equations is not significant and can be omitted.

We have from Eqs. (2.1) – (2.3)

$$\frac{\partial \rho}{\partial t} + \frac{\partial}{\partial x}\left\{\rho v_0 - \tau^{(0)}\left[\frac{\partial}{\partial t}(\rho v_0) + \frac{\partial}{\partial x}\left(\rho v_0^2\right) + \rho \frac{\partial U}{\partial x}\right]\right\} = 0, \qquad (2.4)$$

$$\frac{\partial}{\partial t}(\rho v_0) + \rho\frac{\partial U}{\partial x} + \frac{\partial}{\partial x}\left\{\rho v_0^2 - \tau^{(0)}\left[\frac{\partial}{\partial t}\left(\rho v_0^2\right) + \frac{\partial}{\partial x}\left(\rho v_0^3\right) + 2\rho v_0\frac{\partial U}{\partial x}\right]\right\} = 0, \qquad (2.5)$$

$$\frac{\partial}{\partial t}\left(\rho v_0^2\right) + \frac{\partial}{\partial x}\left\{\rho v_0^3 - \tau^{(0)}\left[\frac{\partial}{\partial t}\left(\rho v_0^3\right) + \frac{\partial}{\partial x}\left(\rho v_0^4\right) + 3\frac{\partial U}{\partial x}\left(\rho v_0^2\right)\right]\right\} +$$

$$+ 2\frac{\partial U}{\partial x}\left\{\rho v_0 - \tau^{(0)}\left[\frac{\partial}{\partial t}(\rho v_0) + \frac{\partial}{\partial x}\left(\rho v_0^2\right) + \rho\frac{\partial U}{\partial x}\right]\right\} = 0 \qquad (2.6)$$

c) Energy equation can be removed from the consideration. It can be realized in the Schroedinger – Madelung model, if two conditions are imposed on the energy equation

$$\rho v_0^3 - \tau^{(0)}\left[\frac{\partial}{\partial t}\left(\rho v_0^3\right) + \frac{\partial}{\partial x}\left(\rho v_0^4\right) + 3\frac{\partial U}{\partial x}\left(\rho v_0^2\right)\right] = C_1, \qquad (2.7)$$

$$\rho v_0 - \tau^{(0)}\left[\frac{\partial}{\partial t}(\rho v_0) + \frac{\partial}{\partial x}\left(\rho v_0^2\right) + \rho\frac{\partial U}{\partial x}\right] = C_2. \qquad (2.8)$$

In this case from energy equation (2.6) follows

$$\frac{\partial}{\partial t}\left(\rho v_0^2\right) = 0, \qquad (2.9)$$

or

$$\rho v_0^2 = F(x). \qquad (2.10)$$

The condition (2.10) does not lead from the limit of approximation. From the law of the energy conservation follows that $C_1 = C_2 = 0$ and both conditions (2.7) and (2.8) are equivalent. Really after multiplication of (2.8) term by term by $v_0^2$ and equalizing the corresponding terms in (2.7), (2.8), we have

$$\frac{\partial}{\partial t}(\rho v_0^3) + \frac{\partial}{\partial x}(\rho v_0^4) + 3\frac{\partial U}{\partial x}(\rho v_0^2) = v_0^2 \frac{\partial}{\partial t}(\rho v_0) + v_0^2 \frac{\partial}{\partial x}(\rho v_0^2) + \rho v_0^2 \frac{\partial U}{\partial x}. \qquad (2.11)$$

But from Eq. (2.11) follows

$$\frac{\partial}{\partial t}\left(\frac{v_0^2}{2}\right) + v_0 \frac{\partial}{\partial x}\left(\frac{v_0^2}{2} + U\right) = 0. \qquad (2.12)$$

For the potential $U$, which does not depend on time, Eq. (2.12) can be rewritten as

$$\frac{\partial}{\partial t}\left(\frac{v_0^2}{2} + U\right) + v_0 \frac{\partial}{\partial x}\left(\frac{v_0^2}{2} + U\right) = 0. \qquad (2.13)$$

Eq. (2.13) is the energy conservation law

$$\frac{v_0^2}{2} + U = const. \qquad (2.14)$$

This result was obtained under the condition $\tau^{(0)} = const$, let us consider now the condition

$$\tau^{(0)} = \tau^{(qu)} = \frac{\hbar}{4m}\frac{1}{v_0^2}, \qquad (2.15)$$

which follows from the "time-energy" uncertainty relation of Heisenberg [2]. From the point of view of non-local physics is obvious that by increasing of particles velocity $v_0$ the non-locality effect should go to decrease; this effect also does not depend on the velocity direction. As result the non-locality effect inversely proportional to velocity squared or in other words – to the energy of particles motion. Instead (2.7), (2.8) we have

$$\rho v_0^3 - \frac{A}{v_0^2}\left[\frac{\partial}{\partial t}(\rho v_0^3) + \frac{\partial}{\partial x}(\rho v_0^4) + 3\frac{\partial U}{\partial x}(\rho v_0^2)\right] = C_1, \qquad (2.16)$$

$$\rho v_0 - \frac{A}{v_0^2}\left[\frac{\partial}{\partial t}(\rho v_0) + \frac{\partial}{\partial x}(\rho v_0^2) + \rho \frac{\partial U}{\partial x}\right] = C_2. \qquad (2.17)$$

After multiplication of (2.17) by $v_0^2$ and equalizing to the expression $\rho v_0^3$ one obtains for $C_1 = C_2 = 0$:

$$v_0^2\left[\frac{\partial}{\partial t}(\rho v_0) + \frac{\partial}{\partial x}(\rho v_0^2) + \rho \frac{\partial U}{\partial x}\right] = \frac{\partial}{\partial t}(\rho v_0^3) + \frac{\partial}{\partial x}(\rho v_0^4) + 3\frac{\partial U}{\partial x}(\rho v_0^2). \qquad (2.18)$$

As we see the condition (2.18) is coincide with the condition (2.11).

Let us suppose that $\tau^{(0)} = const$ and introduce the notation $v_0 = u$ and the following system of scales:

$$\rho_0, \ u_0, \ t_0 = \tau^{(0)}, x_0 = u_0 t_0, U_0 = u_0^2.$$

The dimensionless system of equations takes place:

$$\frac{\partial \tilde{\rho}}{\partial \tilde{t}} + \frac{\partial}{\partial \tilde{x}}\left\{\tilde{\rho}\tilde{u} - \left[\frac{\partial}{\partial \tilde{t}}(\tilde{\rho}\tilde{u}) + \frac{\partial}{\partial \tilde{x}}(\tilde{\rho}\tilde{u}^2) + \tilde{\rho}\frac{\partial \tilde{U}}{\partial \tilde{x}}\right]\right\} = 0, \qquad (2.19)$$

$$\frac{\partial}{\partial \tilde{t}}(\tilde{\rho}\tilde{u}) + \tilde{\rho}\frac{\partial \tilde{U}}{\partial \tilde{x}} + \frac{\partial}{\partial \tilde{x}}\left\{\tilde{\rho}\tilde{u}^2 - \left[\frac{\partial}{\partial \tilde{t}}(\tilde{\rho}\tilde{u}^2) + \frac{\partial}{\partial \tilde{x}}(\tilde{\rho}\tilde{u}^3) + 2\tilde{\rho}\tilde{u}\frac{\partial \tilde{U}}{\partial \tilde{x}}\right]\right\} = 0, \qquad (2.20)$$

where dimensionless symbols are marked by tildes.

The condition (2.9) leads to simplification of the motion equation, one uses also momentum equation for the exclusion of the time derivative in the curl brackets of Eq. (2.20). As result we obtain the system of two dimensionless equations:

$$\frac{\partial \tilde{\rho}}{\partial \tilde{t}} + \frac{\partial}{\partial \tilde{x}}\left\{\tilde{\rho}\tilde{u} - \frac{\partial^2}{\partial \tilde{x}^2}(\tilde{\rho}\tilde{u}^3) - 2\frac{\partial}{\partial \tilde{x}}\left(\tilde{\rho}\tilde{u}\frac{\partial \tilde{U}}{\partial \tilde{x}}\right)\right\} = 0. \qquad (2.21)$$

$$\frac{\partial}{\partial \tilde{t}}(\tilde{\rho}\tilde{u}) + \tilde{\rho}\frac{\partial \tilde{U}}{\partial \tilde{x}} + \frac{\partial}{\partial \tilde{x}}\left\{\tilde{\rho}\tilde{u}^2 - \frac{\partial}{\partial \tilde{x}}(\tilde{\rho}\tilde{u}^3) - 2\tilde{\rho}\tilde{u}\frac{\partial \tilde{U}}{\partial \tilde{x}}\right\} = 0, \qquad (2.22)$$

### 3. Wave solutions under condition $\tau^{(0)} = const$ and without dissipation.

With the aim to find wave solutions of Eqs. (2.21), (2.22) we use the moving coordinate system, where

$$\tilde{\xi} = \tilde{x} - \tilde{C}\tilde{t}. \qquad (3.1)$$

In moving coordinate system all dependent hydrodynamic values are function of $(\tilde{\xi}, \tilde{t})$ and equations (2.21), (2.22) take the form

$$\frac{\partial \tilde{\rho}}{\partial \tilde{t}} - \tilde{C}\frac{\partial \tilde{\rho}}{\partial \tilde{\xi}} + \frac{\partial}{\partial \tilde{\xi}}\left\{\tilde{\rho}\tilde{u} - \frac{\partial^2}{\partial \tilde{\xi}^2}(\tilde{\rho}\tilde{u}^3) - 2\frac{\partial}{\partial \tilde{\xi}}\left(\tilde{\rho}\tilde{u}\frac{\partial \tilde{U}}{\partial \tilde{\xi}}\right)\right\} = 0, \qquad (3.2)$$

$$\frac{\partial}{\partial \tilde{t}}(\tilde{\rho}\tilde{u}) - \tilde{C}\frac{\partial}{\partial \tilde{\xi}}(\tilde{\rho}\tilde{u}) + \tilde{\rho}\frac{\partial \tilde{U}}{\partial \tilde{\xi}} + \frac{\partial}{\partial \tilde{\xi}}\left\{\tilde{\rho}\tilde{u}^2 - \frac{\partial}{\partial \tilde{\xi}}(\tilde{\rho}\tilde{u}^3) - 2\tilde{\rho}\tilde{u}\frac{\partial \tilde{U}}{\partial \tilde{\xi}}\right\} = 0, \qquad (3.3)$$

For the solution there is no explicit dependence on time for coordinate system moving with the phase velocity $\tilde{C}$. In moving coordinate system the soliton either fixed or all points of the single wave are moving with the same velocity. We have from (3.2), (3.3)

$$-\tilde{C}\frac{\partial \tilde{\rho}}{\partial \tilde{\xi}} + \frac{\partial}{\partial \tilde{\xi}}\left\{\tilde{\rho}\tilde{u} - \frac{\partial^2}{\partial \tilde{\xi}^2}(\tilde{\rho}\tilde{u}^3) - 2\frac{\partial}{\partial \tilde{\xi}}\left(\tilde{\rho}\tilde{u}\frac{\partial \tilde{U}}{\partial \tilde{\xi}}\right)\right\} = 0, \qquad (3.4)$$

$$-\tilde{C}\frac{\partial}{\partial \tilde{\xi}}(\tilde{\rho}\tilde{u}) + \tilde{\rho}\frac{\partial \tilde{U}}{\partial \tilde{\xi}} + \frac{\partial}{\partial \tilde{\xi}}\left\{\tilde{\rho}\tilde{u}^2 - \frac{\partial}{\partial \tilde{\xi}}(\tilde{\rho}\tilde{u}^3) - 2\tilde{\rho}\tilde{u}\frac{\partial \tilde{U}}{\partial \tilde{\xi}}\right\} = 0, \qquad (3.5)$$

After integration of (3.4):

$$\tilde{\rho}(\tilde{u} - \tilde{C}) = \frac{\partial}{\partial \tilde{\xi}}\left\{\tilde{u}^2\frac{\partial}{\partial \tilde{\xi}}(\tilde{\rho}\tilde{u}) + \tilde{\rho}\tilde{u}\frac{\partial}{\partial \tilde{\xi}}(\tilde{u}^2 + 2\tilde{U})\right\} - \tilde{C}_1. \qquad (3.6)$$

Let us use the law of the energy conservation in the form $\frac{\tilde{u}^2}{2} + \tilde{U} = \tilde{H} + \frac{\tilde{u}_\phi^2}{2}$, where phase velocity $u_\phi = C$ is introduced. Then from (3.6)

$$\tilde{\rho}(\tilde{u} - \tilde{C}) = 2\frac{\partial}{\partial\tilde{\xi}}\left\{[\tilde{H} - \tilde{U}]\frac{\partial}{\partial\tilde{\xi}}(\tilde{\rho}\tilde{u})\right\} - \tilde{C}_1. \qquad (3.7)$$

But by $\tilde{u} = \tilde{u}_\phi = \tilde{C}$ we have $\tilde{U} = \tilde{H}$, therefore $\tilde{C}_1 = 0$. Let us introduce the dependent variable $\tilde{y} = \tilde{\rho}\tilde{u}$ and transform (3.5), (3.7) using the condition

$$\tilde{u}^2 = 2(\tilde{H}_{en} - \tilde{U}), \text{ where } \tilde{H}_{en} = \tilde{H} + \frac{\tilde{u}_\phi^2}{2}.$$

We have

$$-\tilde{C}\frac{\partial\tilde{y}}{\partial\tilde{\xi}} + \frac{1}{\tilde{C}}\frac{\partial\tilde{U}}{\partial\tilde{\xi}}\left\{\tilde{y} - 2\frac{\partial}{\partial\tilde{\xi}}\left[(\tilde{H}_{en} - \tilde{U})\frac{\partial\tilde{y}}{\partial\tilde{\xi}}\right]\right\} +$$

$$+ \frac{\partial}{\partial\tilde{\xi}}\left\{\frac{2}{\tilde{C}}\left[\tilde{y} - 2\frac{\partial}{\partial\tilde{\xi}}\left[(\tilde{H}_{en} - \tilde{U})\frac{\partial\tilde{y}}{\partial\tilde{\xi}}\right]\right](\tilde{H}_{en} - \tilde{U}) - 2(\tilde{H}_{en} - \tilde{U})\frac{\partial\tilde{y}}{\partial\tilde{\xi}}\right\} = 0. \qquad (3.8)$$

Let us consider now the example of Eq. (3.8) solution for the case of the oscillator motion

$$\tilde{U} = \tilde{\xi}^2, \tilde{H}_{en} - \tilde{U} = 1 - \tilde{\xi}^2; \tilde{C} = 1. \qquad (3.9)$$

The equality $\tilde{C} = 1$ means that the velocity of moving coordinate system is equal to the velocity scale $u_0$. After introducing new dependent variable $\tilde{y} = \tilde{\rho}\tilde{u}$ we have from (3.7) - (3.9):

$$2(1 - \tilde{\xi}^2)\left\{2(1 - \tilde{\xi}^2)\frac{\partial^3\tilde{y}}{\partial\tilde{\xi}^3} + (1 - 10\tilde{\xi})\frac{\partial^2\tilde{y}}{\partial\tilde{\xi}^2} - 5\frac{\partial\tilde{y}}{\partial\tilde{\xi}}\right\} +$$

$$+ (1 + 8\tilde{\xi}^2 - 4\tilde{\xi})\frac{\partial\tilde{y}}{\partial\tilde{\xi}} + 2\tilde{\xi}\tilde{y} = 0, \qquad (3.10)$$

$$\frac{\partial\tilde{\rho}}{\partial\tilde{\xi}} = -2(1 - \tilde{\xi}^2)\frac{\partial^3\tilde{y}}{\partial\tilde{\xi}^3} + 8\tilde{\xi}\frac{\partial^2\tilde{y}}{\partial\tilde{\xi}^2} + 5\frac{\partial\tilde{y}}{\partial\tilde{\xi}}. \qquad (3.11)$$

Figures 1 and 2 reflect some results of calculations realized according formulae (3.10), (3.11) with the help of Maple 9. The following notations on figures are used: $r$ - density $\tilde{\rho}$ (solid line), $u$ - momentum $\tilde{\rho}\tilde{u}$ (dashed line), kinetic energy $w = \tilde{\rho}\tilde{u}^2/2$ (dash dotted line). Explanations placed under all following figures, contain Maple's notations. For example, expressions $D(u)(0) = 0$, $D(D(u))(0) = 0$ mean in usual notations $\frac{\partial\tilde{u}}{\partial\tilde{\xi}}(0) = 0, \frac{\partial^2\tilde{u}}{\partial\tilde{\xi}^2}(0) = 0$. Solutions look like solitons with practically constant velocity; the waves placed between singular domains, derivatives of $\tilde{\rho}$ take the discrete constant values.

```
Warning, cannot evaluate the solution further right of 1.2000000, probably a
singularity
Warning, cannot evaluate the solution further left of -1.2000000, probably a
singularity
```

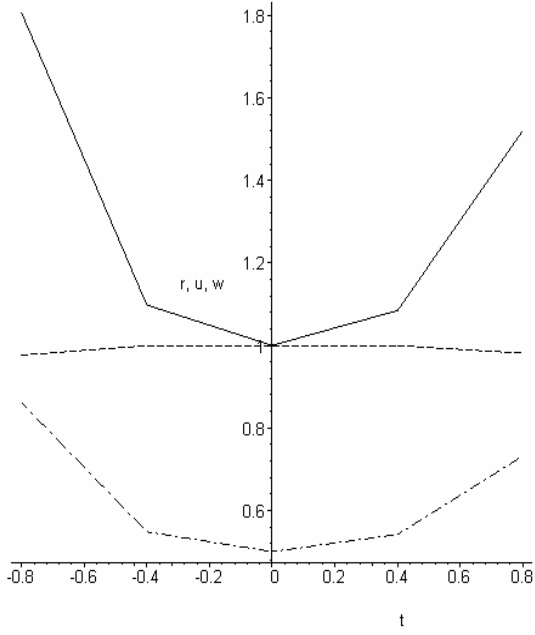 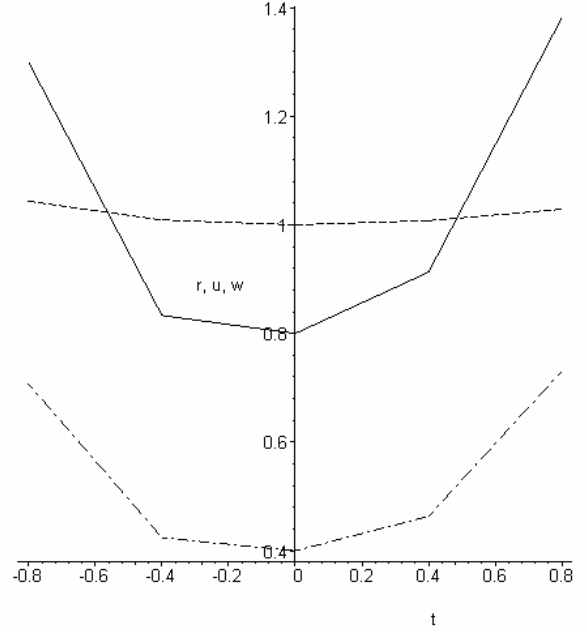

Fig. 1.　　　　　　　　　　　　　　　Fig. 2.

```
u(0)=1,D(u)(0)=0,D(D(u))(0)=0,        u(0)=1,D(u)(0)=0,D(D(u))(0)=.1,
      r(0)=1,w(0)=0.5.                     r(0)=0.8,w(0)=0.4
```

Let us consider another particular case – two displaced oscillators; $\tau^{(0)} = const$; no dissipation. We suppose

$$\tilde{U} = \tilde{\xi}^2 + (\tilde{\xi} - 1)^2, \ \tilde{C} = 1, \ \tilde{H}_{en} - \tilde{U} = 2\tilde{\xi}(1 - \tilde{\xi}).$$

Equations (3.7), (3.8) transform into the set

$$\frac{\partial \tilde{\rho}}{\partial \tilde{\xi}} = -4\tilde{\xi}(1-\tilde{\xi})\frac{\partial^3 \tilde{y}}{\partial \tilde{\xi}^3} - 8(1-2\tilde{\xi})\frac{\partial^2 \tilde{y}}{\partial \tilde{\xi}^2} + 9\frac{\partial \tilde{y}}{\partial \tilde{\xi}}. \tag{3.12}$$

$$16\tilde{\xi}^2(1-\tilde{\xi})^2 \frac{\partial^3 y}{\partial \tilde{\xi}^3} + \tilde{\xi}(1-\tilde{\xi})(44-80\tilde{\xi})\frac{\partial^2 y}{\partial \tilde{\xi}^2} + (68\tilde{\xi}^2 - 76\tilde{\xi} + 13)\frac{\partial \tilde{y}}{\partial \tilde{\xi}} -$$
$$- (2-4\tilde{\xi})\tilde{y} = 0, \tag{3.13}$$

Figures 3 and 4 reflect some results of calculations realized according formulae (3.12), (3.13). As before the following notations on figures are used: $r$ - density $\tilde{\rho}$ (solid line), $u$ - momentum $\tilde{\rho}\tilde{u}$ (dashed line), kinetic energy $w = \tilde{\rho}\tilde{u}^2/2$ (dash dotted line).

```
Warning, cannot evaluate the solution further right of 1.3000000, probably a
singularity
Warning, cannot evaluate the solution further left of -.30000000, probably a
singularity
```

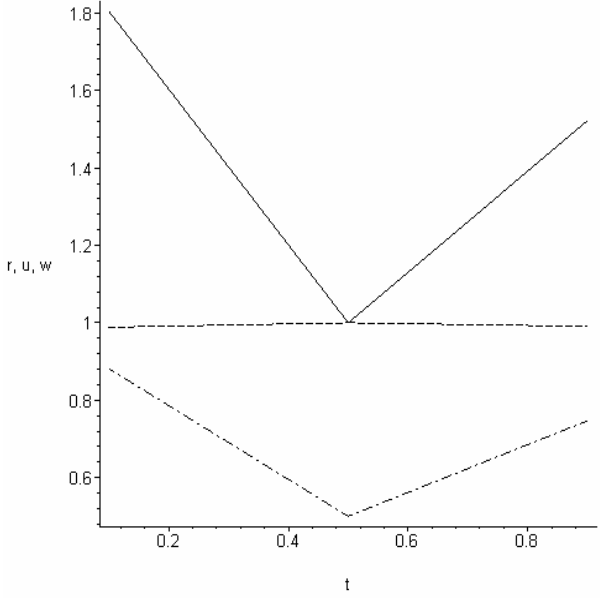
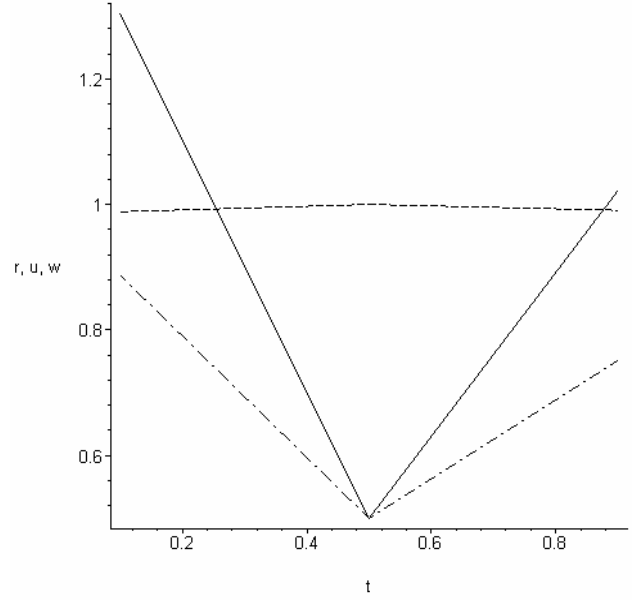

Fig. 3.                         Fig. 4.

```
u(0.5)=1,D(u)(0.5)=0,D(D(u))(0.5)=0,         u(.5)=1,D(u)(.5)=0,
r(0.5)=1,w(0.5)=0.5                          D(D(u))(.5)=0,r(0.5)=.5,
                                             w(0.5)=0.5
```

**4. Derivation of basic equations with taking into account of the uncertainty principle but without dissipation.**

On our following way to the numerical solution of the full system of equation we consider the important particular case when the energy equation can be excluded from the consideration. The condition (2.9) leads to simplification of the momentum equation and then to the next system of dimensionless equations (see also (2.4), (2.5))

$$\frac{\partial \tilde{\rho}}{\partial \tilde{t}} + \frac{\partial}{\partial \tilde{x}}\left\{\tilde{\rho}\tilde{u} - \frac{\tilde{A}}{\tilde{u}^2}\left[\frac{\partial}{\partial \tilde{t}}(\tilde{\rho}\tilde{u}) + \frac{\partial}{\partial \tilde{x}}(\tilde{\rho}\tilde{u}^2) + \tilde{\rho}\frac{\partial \tilde{U}}{\partial \tilde{x}}\right]\right\} = 0, \qquad (4.1)$$

$$\frac{\partial}{\partial \tilde{t}}(\tilde{\rho}\tilde{u}) + \tilde{\rho}\frac{\partial \tilde{U}}{\partial \tilde{x}} + \frac{\partial}{\partial \tilde{x}}\left\{\tilde{\rho}\tilde{u}^2 - \frac{\tilde{A}}{\tilde{u}^2}\left[\frac{\partial}{\partial \tilde{x}}(\tilde{\rho}\tilde{u}^3) + 2\tilde{\rho}\tilde{u}\frac{\partial \tilde{U}}{\partial \tilde{x}}\right]\right\} = 0. \qquad (4.2)$$

Consider the scales system for this case and remind the condition (2.15) which follows from the "time-energy" uncertainty relation of Heisenberg [2]. Now $\tau^{(0)} = \tau^{(qu)} = \frac{\hbar}{4m}\frac{1}{u^2}$ and we use the following system of scales.

$$\rho_0,\ u_0, x_0 = u_0 t_0, U_0 = u_0^2,\ t_0 = \frac{\hbar}{4m}\frac{1}{u_0^2}.$$

In this case $\tau^{(0)} = \tau^{(qu)} = t_0 \frac{u_0^2}{u^2} = t_0 \frac{1}{\tilde{u}^2}$. For this choice of scales $\tilde{A} = 1$. With the aim to find wave solutions of Eqs. (4.1), (4.2) we use the moving coordinate system, where $\tilde{\xi} = \tilde{x} - \tilde{C}\tilde{t}$. In moving coordinate system all dependent hydrodynamic values are function of $(\tilde{\xi}, \tilde{t})$ and equations (4.1), (4.2) take the form

$$\frac{\partial \tilde{\rho}}{\partial \tilde{t}} - \tilde{C} \frac{\partial \tilde{\rho}}{\partial \tilde{\xi}} + \frac{\partial}{\partial \tilde{\xi}} \left\{ \tilde{\rho}\tilde{u} - \frac{1}{\tilde{u}^2} \left[ \frac{\partial}{\partial \tilde{t}} (\tilde{\rho}\tilde{u}) - \tilde{C} \frac{\partial}{\partial \tilde{\xi}} (\tilde{\rho}\tilde{u}) + \frac{\partial}{\partial \tilde{\xi}} (\tilde{\rho}\tilde{u}^2) + \rho \frac{\partial \tilde{U}}{\partial \tilde{\xi}} \right] \right\} = 0, \quad (4.3)$$

$$\frac{\partial}{\partial \tilde{t}} (\tilde{\rho}\tilde{u}) - \tilde{C} \frac{\partial}{\partial \tilde{\xi}} (\tilde{\rho}\tilde{u}) + \tilde{\rho} \frac{\partial \tilde{U}}{\partial \tilde{\xi}} + \frac{\partial}{\partial \tilde{\xi}} \left\{ \tilde{\rho}\tilde{u}^2 - \frac{1}{\tilde{u}^2} \left[ \frac{\partial}{\partial \tilde{\xi}} (\tilde{\rho}\tilde{u}^3) + 2\tilde{\rho}\tilde{u} \frac{\partial \tilde{U}}{\partial \tilde{\xi}} \right] \right\} = 0, \quad (4.4)$$

As before we state that in moving coordinate system the soliton either fixed or all points of the single wave are moving with the same velocity, there is no explicit dependence on time. We assume also that $\tilde{C} = 1$. The system of dimensionless equations should be solved

$$- \frac{\partial \tilde{\rho}}{\partial \tilde{\xi}} + \frac{\partial}{\partial \tilde{\xi}} \left\{ \tilde{\rho}\tilde{u} - \frac{1}{\tilde{u}^2} \left[ - \frac{\partial}{\partial \tilde{\xi}} (\tilde{\rho}\tilde{u}) + \frac{\partial}{\partial \tilde{\xi}} (\tilde{\rho}\tilde{u}^2) + \rho \frac{\partial \tilde{U}}{\partial \tilde{\xi}} \right] \right\} = 0, \quad (4.5)$$

$$- \frac{\partial}{\partial \tilde{\xi}} (\tilde{\rho}\tilde{u}) + \tilde{\rho} \frac{\partial \tilde{U}}{\partial \tilde{\xi}} + \frac{\partial}{\partial \tilde{\xi}} \left\{ \tilde{\rho}\tilde{u}^2 - \frac{1}{\tilde{u}^2} \left[ \frac{\partial}{\partial \tilde{\xi}} (\tilde{\rho}\tilde{u}^3) + 2\tilde{\rho}\tilde{u} \frac{\partial \tilde{U}}{\partial \tilde{\xi}} \right] \right\} = 0, \quad (4.6)$$

Figures 5 and 6 reflect some results of calculations realized according formulae (4.5), (4.6) for oscillator ($\tilde{U} = \tilde{\xi}^2$) with the help of Maple 9. The following notations on figures are used: $r$ - density $\tilde{\rho}$ (solid line), $u$ - velocity $\tilde{u}$ (dashed line), momentum - $\tilde{\rho}\tilde{u}$ (dash dot line). Independent dimensionless variable $t$ on graphs is $\tilde{\xi}$.

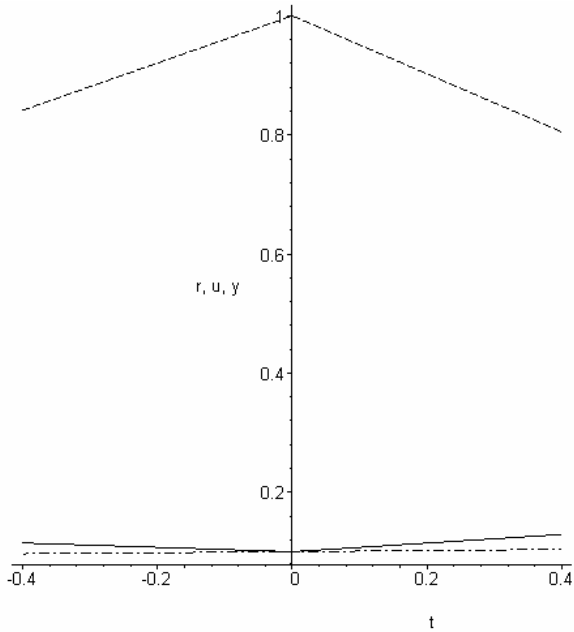
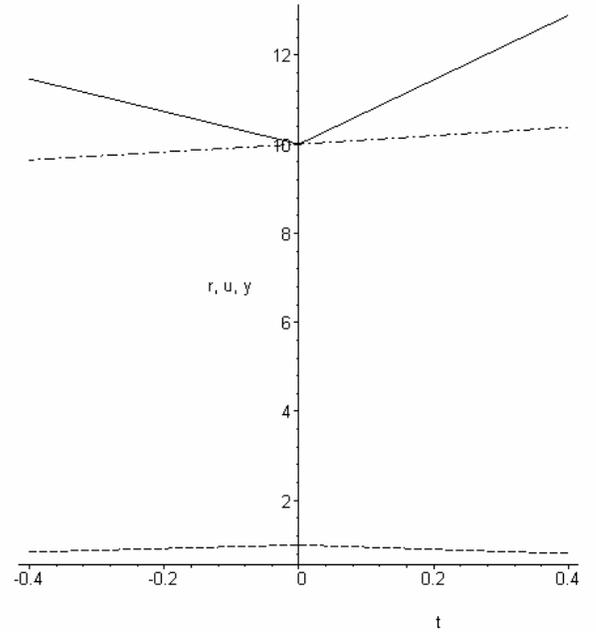

Fig. 5.                                              Fig. 6.

```
u(0)=1, D(u)(0)=0, r(0)=.1,        u(0)=1, D(u)(0)=0, r(0)=10,
D(r)(0)=0,y(0)=0.1.                  D(r)(0)=0, y(0)=10.
```

```
Warning, cannot evaluate the solution further right of .80000000, probably a
singularity. Warning, cannot evaluate the solution further left of -.80000000,
probably a singularity
```

## 5. Solution of the full system of equations for the case $\tau^{(0)} = const$.

Using the developed theory of soliton's solution we obtain from the system (2.1) – (2.3) dimensionless equations in moving coordinate system ($\tilde{C} = 1$):

(continuity equation)

$$\frac{\partial^2 \tilde{\rho}}{\partial \tilde{\xi}^2} + \frac{\partial \tilde{\rho}}{\partial \tilde{\xi}} - 2\frac{\partial^2}{\partial \tilde{\xi}^2}(\tilde{\rho}\tilde{u}) - \frac{\partial}{\partial \tilde{\xi}}(\tilde{\rho}\tilde{u}) + \frac{\partial^2}{\partial \tilde{\xi}^2}(\tilde{\rho}\tilde{u}^2) + \frac{\partial^2 \tilde{p}}{\partial \tilde{\xi}^2} + \frac{\partial}{\partial \tilde{\xi}}\left(\tilde{\rho}\frac{\partial \tilde{U}}{\partial \tilde{\xi}}\right) = 0, \quad (5.1)$$

(momentum equation)

$$\frac{\partial^2}{\partial \tilde{\xi}^2}(\tilde{\rho}\tilde{u}) + \frac{\partial}{\partial \tilde{\xi}}(\tilde{\rho}\tilde{u}) - 2\frac{\partial^2}{\partial \tilde{\xi}^2}(\tilde{\rho}\tilde{u}^2) - 2\frac{\partial^2 \tilde{p}}{\partial \tilde{\xi}^2} - \frac{\partial}{\partial \tilde{\xi}}(\tilde{\rho}\tilde{u}^2) - \frac{\partial \tilde{p}}{\partial \tilde{\xi}} + \frac{\partial^2}{\partial \tilde{\xi}^2}(\tilde{\rho}\tilde{u}^3) +$$

$$+ 3\frac{\partial^2}{\partial \tilde{\xi}^2}(\tilde{p}\tilde{u}) + 2\frac{\partial}{\partial \tilde{\xi}}\left(\tilde{\rho}\tilde{u}\frac{\partial \tilde{U}}{\partial \tilde{\xi}}\right) - \frac{\partial}{\partial \tilde{\xi}}\left(\tilde{\rho}\frac{\partial \tilde{U}}{\partial \tilde{\xi}}\right) - \frac{\partial \tilde{U}}{\partial \tilde{\xi}}\left(\tilde{\rho} + \frac{\partial \tilde{\rho}}{\partial \tilde{\xi}} - \frac{\partial}{\partial \tilde{\xi}}(\tilde{\rho}\tilde{u})\right) = 0, \quad (5.2)$$

(energy equation)

$$3\frac{\partial^2 \tilde{p}}{\partial \tilde{\xi}^2} + \frac{\partial}{\partial \tilde{\xi}}(\tilde{\rho}\tilde{u}^2) + 3\frac{\partial \tilde{p}}{\partial \tilde{\xi}} + \frac{\partial^2}{\partial \tilde{\xi}^2}(\tilde{\rho}\tilde{u}^2) - 2\frac{\partial^2}{\partial \tilde{\xi}^2}(\tilde{\rho}\tilde{u}^3) - 10\frac{\partial^2}{\partial \tilde{\xi}^2}(\tilde{p}\tilde{u}) -$$

$$- 2\frac{\partial}{\partial \tilde{\xi}}\left(\tilde{\rho}\tilde{u}\frac{\partial \tilde{U}}{\partial \tilde{\xi}}\right) - \frac{\partial}{\partial \tilde{\xi}}(\tilde{\rho}\tilde{u}^3) - 5\frac{\partial}{\partial \tilde{\xi}}(\tilde{p}\tilde{u}) + \frac{\partial^2}{\partial \tilde{\xi}^2}(\tilde{\rho}\tilde{u}^4) + 8\frac{\partial^2}{\partial \tilde{\xi}^2}(\tilde{p}\tilde{u}^2) + 5\frac{\partial^2}{\partial \tilde{\xi}^2}\frac{\tilde{p}^2}{\tilde{\rho}} +$$

$$+ \frac{\partial}{\partial \tilde{\xi}}\left[\frac{\partial \tilde{U}}{\partial \tilde{\xi}}(5\tilde{p} + 3\tilde{\rho}\tilde{u}^2)\right] - 2\frac{\partial \tilde{U}}{\partial \tilde{\xi}}\left(\tilde{\rho}\tilde{u} + \frac{\partial}{\partial \tilde{\xi}}(\tilde{\rho}\tilde{u}) - \frac{\partial}{\partial \tilde{\xi}}(\tilde{\rho}\tilde{u}^2) - \frac{\partial \tilde{p}}{\partial \tilde{\xi}} - \tilde{\rho}\frac{\partial \tilde{U}}{\partial \tilde{\xi}}\right) = 0,$$

(5.3)

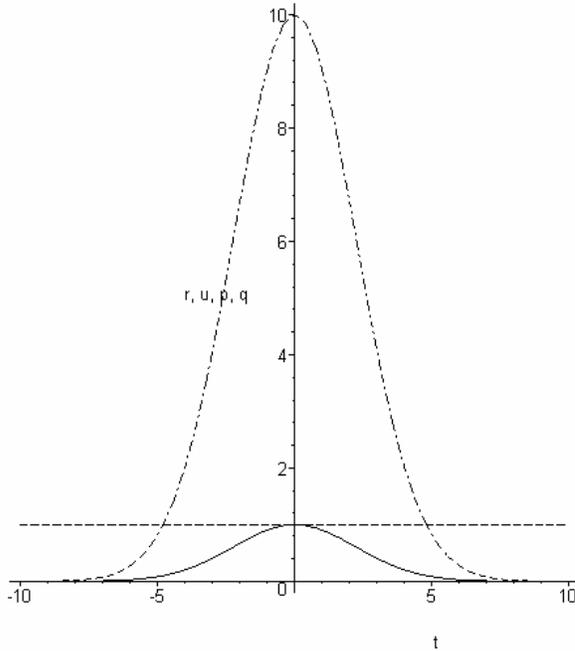
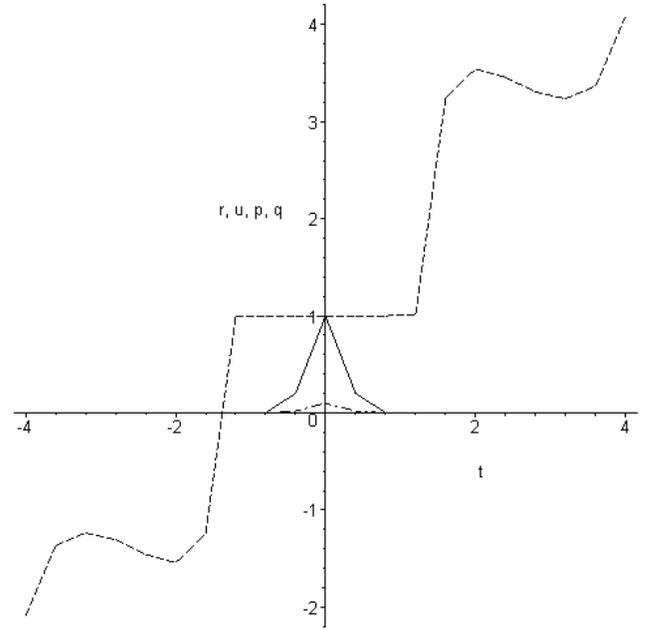

Fig. 7.            Fig. 8.

Figures 7 and 8 reflect some results of calculations realized according formulae (4.5), (4.6) for oscillator ($\tilde{U} = \tilde{\xi}^2$) with the help of Maple 9. The following notations on figures are used: $r$ - density $\tilde{\rho}$ (solid line), $u$ - velocity $\tilde{u}$ (dashed line), $p$ - pressure $\tilde{p}$ (dash dot line), $q$ - momentum $\tilde{\rho}\tilde{u}$ (dotted line). Independent dimensionless variable $t$ on graphs is $\tilde{\xi}$. Numerical results reflected by Fig. 5 and 6 obtained for initial perturbations
**r(0)=1,D(r)(0)=0,u(0)=1,D(u)(0)=0,p(0)=10,D(p)(0)=0,q(0)=1,(Fig.7);**
**r(0)=1,D(r)(0)=0,u(0)=1,D(u)(0)=0,p(0)=.1,D(p)(0)=0,q(0)=1,(Fig.8).**

The results shown in Fig. 7 can be treated as formation of the "ideal" soliton, when all elements of a single wave are moving with the strictly constant velocity and the numerical method does not catch the presence of singular domains. Therefore if in the initial time moment the perturbation (indicated as initial conditions) takes place, then the formatted single wave is expanding along $x$ - axis without dissipation.

Let us consider now another case (two displaced oscillators) of application of Eqs. (5.1) – (5.3), for which $\tilde{U} = \tilde{\xi}^2 + (\tilde{\xi} - 1)^2$. Figures 9 and 10 correspond to the same perturbations:
**r(0)=1,D(r)(0)=0,u(0)=1,D(u)(0)=0,p(0)=1,D(p)(0)=0,q(0)=1, (Fig.9),**
**r(0.5)=1,D(r)(0.5)=0,u(0.5)=1,D(u)(0.5)=0,p(0.5)=1,D(p)(0.5)=0,q(0.5)=1, (Fig.10),** but applied to the different $\tilde{\xi}$ - points. As in the previous case: $r$ - density $\tilde{\rho}$ (solid line), $u$ - velocity $\tilde{u}$ (dashed line), $p$ - pressure $\tilde{p}$ (dash dot line), $q$ - momentum $\tilde{\rho}\tilde{u}$ (dotted line). Independent dimensionless variable $t$ on graphs is $\tilde{\xi}$.

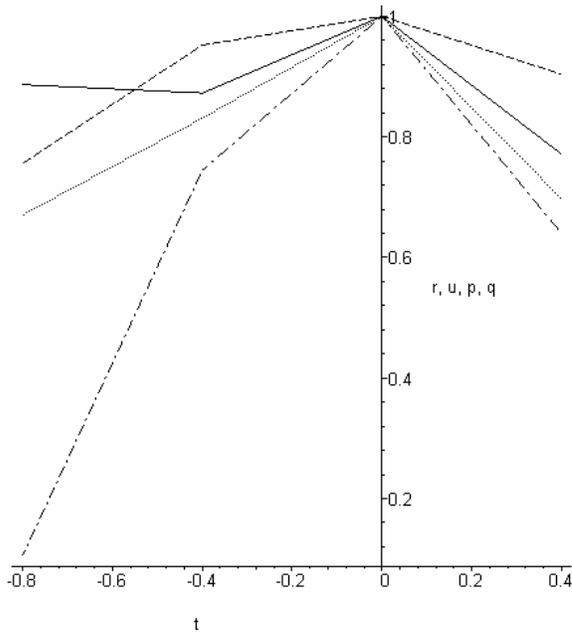
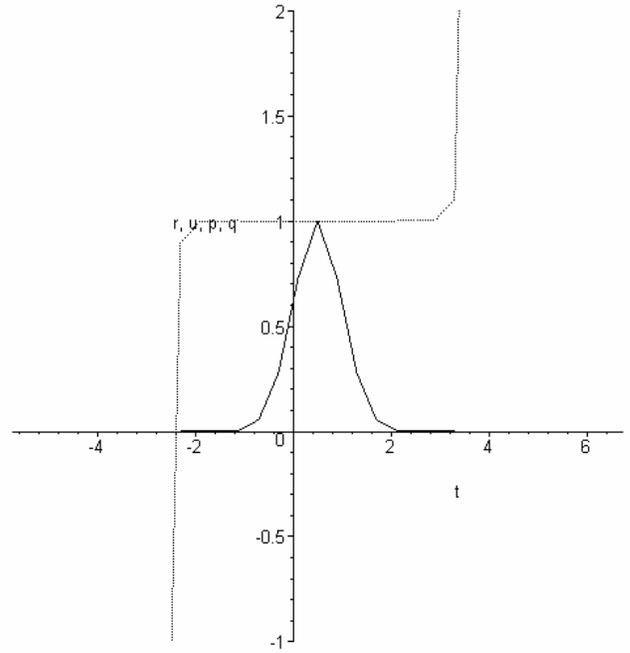

Fig. 9.  Fig. 10.

## 6. Solution of full system of equations with taking into account of the uncertainty principle.

Let us consider the numerical solutions of full system of equations (2.1) – (2.3) written in the dimensionless form for the scales $\rho_0$, $u_0$, $x_0 = u_0 t_0$, $U_0 = u_0^2$, $t_0 = \dfrac{\hbar}{4m}\dfrac{1}{u_0^2}$ and conditions $\tilde{C} = 1$, $\tilde{A} = 1$. The equations take the form

$$\frac{\partial \tilde{\rho}}{\partial \tilde{\xi}} - \frac{\partial \tilde{\rho}\tilde{u}}{\partial \tilde{\xi}} + \frac{\partial}{\partial \tilde{\xi}}\left\{\frac{1}{\tilde{u}^2}\left[\frac{\partial}{\partial \tilde{\xi}}\left(\tilde{p} + \tilde{\rho} + \tilde{\rho}\tilde{u}^2 - 2\tilde{\rho}\tilde{u}\right) + \tilde{\rho}\frac{\partial \tilde{U}}{\partial \tilde{\xi}}\right]\right\} = 0, \quad (6.1)$$

$$\frac{\partial}{\partial \tilde{\xi}}\left(\tilde{\rho}\tilde{u}^2 + \tilde{p} - \tilde{\rho}\tilde{u}\right) + \frac{\partial}{\partial \tilde{\xi}}\left\{\frac{1}{\tilde{u}^2}\left[\frac{\partial}{\partial \tilde{\xi}}\left(2\tilde{\rho}\tilde{u}^2 - \tilde{\rho}\tilde{u} + 2\tilde{p} - \tilde{\rho}\tilde{u}^3 - 3\tilde{p}\tilde{u}\right) + \tilde{\rho}\frac{\partial \tilde{U}}{\partial \tilde{\xi}}\right]\right\} +$$

$$+ \frac{\partial \tilde{U}}{\partial \tilde{\xi}}\left\{\tilde{\rho} - \frac{1}{\tilde{u}^2}\left[-\frac{\partial \tilde{\rho}}{\partial \tilde{\xi}} + \frac{\partial}{\partial \tilde{\xi}}(\tilde{\rho}\tilde{u})\right]\right\} - 2\frac{\partial}{\partial \tilde{\xi}}\left\{\frac{\tilde{\rho}}{\tilde{u}}\frac{\partial \tilde{U}}{\partial \tilde{\xi}}\right\} = 0,$$

(6.2)

$$\frac{\partial}{\partial \tilde{\xi}}\left(\tilde{\rho}\tilde{u}^2 + 3\tilde{p} - \tilde{\rho}\tilde{u}^3 - 5\tilde{p}\tilde{u}\right) -$$

$$- \frac{\partial}{\partial \tilde{\xi}}\left\{\frac{1}{\tilde{u}^2}\frac{\partial}{\partial \tilde{\xi}}\left(2\tilde{\rho}\tilde{u}^3 + 10\tilde{p}\tilde{u} - \tilde{\rho}\tilde{u}^2 - 3\tilde{p} - \tilde{\rho}\tilde{u}^4 - 8\tilde{p}\tilde{u}^2 - 5\frac{\tilde{p}^2}{\tilde{\rho}}\right)\right\} +$$

(6.3)

$$+ \frac{\partial}{\partial \tilde{\xi}}\left\{\frac{1}{\tilde{u}^2}\left(3\tilde{\rho}\tilde{u}^2 + 5\tilde{p}\right)\frac{\partial \tilde{U}}{\partial \tilde{\xi}}\right\} - 2\tilde{\rho}\tilde{u}\frac{\partial \tilde{U}}{\partial \tilde{\xi}} - 2\frac{\partial}{\partial \tilde{\xi}}\left\{\frac{\tilde{\rho}}{\tilde{u}}\frac{\partial \tilde{U}}{\partial \tilde{\xi}}\right\} +$$

$$+ \frac{2}{\tilde{u}^2}\frac{\partial \tilde{U}}{\partial \tilde{\xi}}\left[-\frac{\partial}{\partial \tilde{\xi}}(\tilde{\rho}\tilde{u}) + \frac{\partial}{\partial \tilde{\xi}}(\tilde{\rho}\tilde{u}^2 + \tilde{p}) + \tilde{\rho}\frac{\partial \tilde{U}}{\partial \tilde{\xi}}\right] = 0,$$

We begin here from the case $\tilde{U} = \tilde{\xi}^2$, corresponding to quantum oscillator. Figures 11-14 demonstrate the function evolution by the increasing of dimensionless density from $\tilde{\rho} = r(t) = 0.1$ to $\tilde{\rho} = r(t) = 15$, correspondingly:

```
u(0)=1,p(0)=1,r(0)=.1,D(u)(0)=0,D(p)(0)=0,D(r)(0)=0;
u(0)=1,p(0)=1,r(0)=.5,D(u)(0)=0,D(p)(0)=0,D(r)(0)=0});
u(0)=1,p(0)=1,r(0)=5,D(u)(0)=0,D(p)(0)=0,D(r)(0)=0});
u(0)=1,p(0)=1,r(0)=15,D(u)(0)=0,D(p)(0)=0,D(r)(0)=0});
```

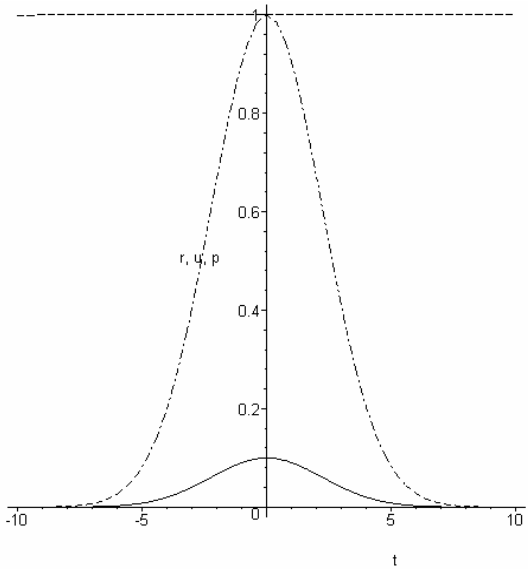 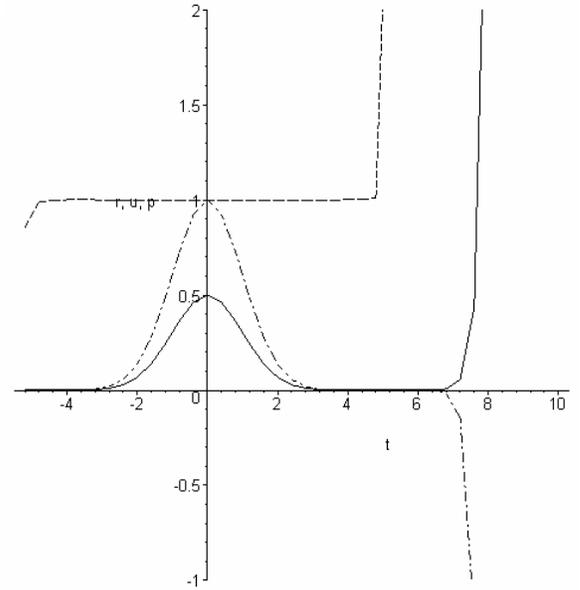

Fig. 11.  Fig. 12.

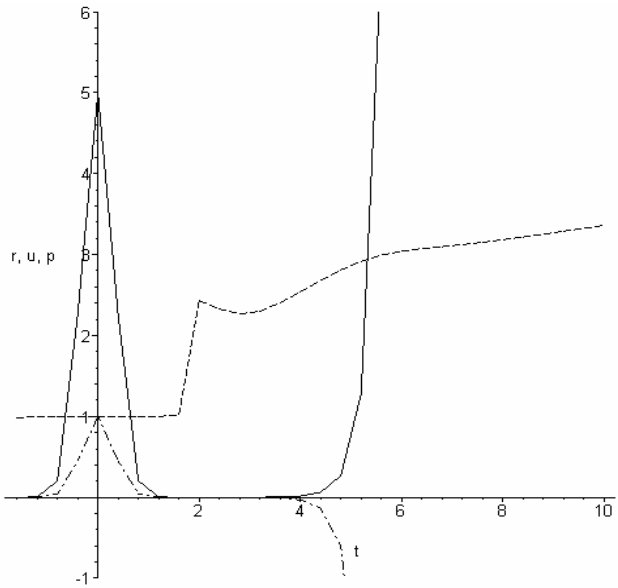 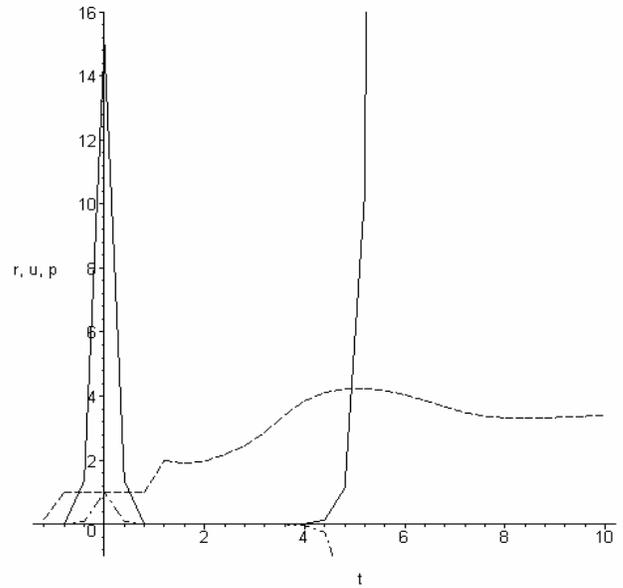

Fig. 13.    Fig. 14.

The following notations on figures 11 – 14 are used: $r$ - density $\tilde{\rho}$ (solid line), $u$ - velocity $\tilde{u}$ (dashed line), $p$ - pressure $\tilde{p}$ (dash dot line). It can be seen from Fig. 11 – 14 approaching of singular domains to solitons as increasing of perturbation of dimensionless density $r \equiv \tilde{\rho}$.

Fig. 15, 16 demonstrate the function evolution by the increasing of dimensionless pressure from $\tilde{p} = p(t) = 10$ to $\tilde{p} = p(t) = 50$, correspondingly:

```
u(0)=1,p(0)=10,r(0)=1,D(u)(0)=0,D(p)(0)=0,D(r)(0)=0});
u(0)=1,p(0)=50,r(0)=1,D(u)(0)=0,D(p)(0)=0,D(r)(0)=0.
```

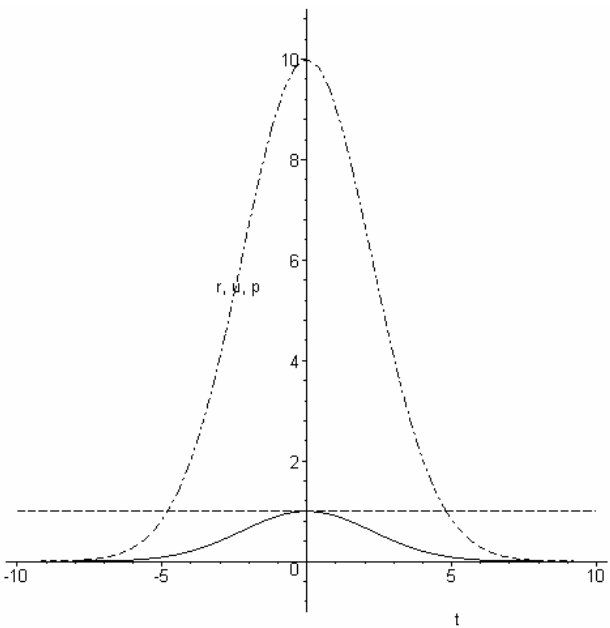 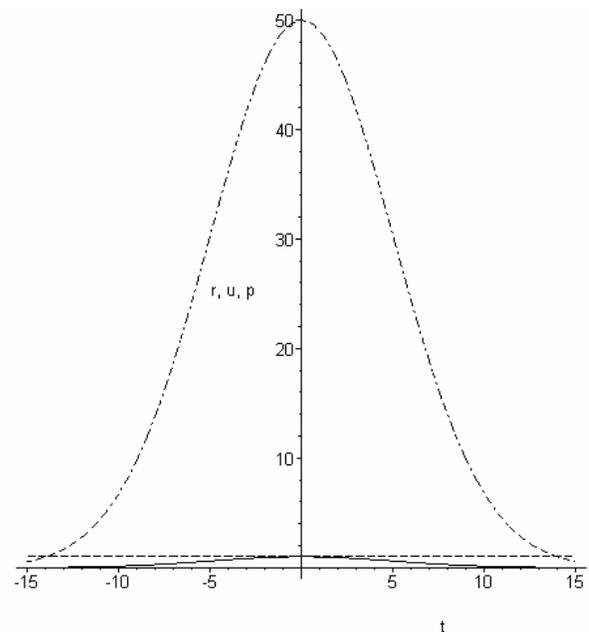

Fig. 15.    Fig. 16.

The following case to be considered is displaced oscillators for which $\tilde{U} = \tilde{\xi}^2 + (\tilde{\xi} - 1)^2$. The numerical solutions of Eqs. (6.1) – (6.3) are reflected on Fig. 17 – 20 for initial conditions placed in the different points $\tilde{\xi}$, correspondingly:

```
u(0)=0.5,p(0)=10,r(0)=1,D(u)(0)=0,D(p)(0)=0,D(r)(0)=0});
u(0.5)=1,p(0.5)=10,r(0.5)=1,D(u)(0.5)=0,D(p)(0.5)=0,D(r)(0.5)=0;
u(0)=1,p(0)=1,r(0)=1,D(u)(0)=0,D(p)(0)=0,D(r)(0)=0;
u(0.5)=1,p(0.5)=1,r(0.5)=1,D(u)(0.5)=0,D(p)(0.5)=0,D(r)(0.5)=0.
```

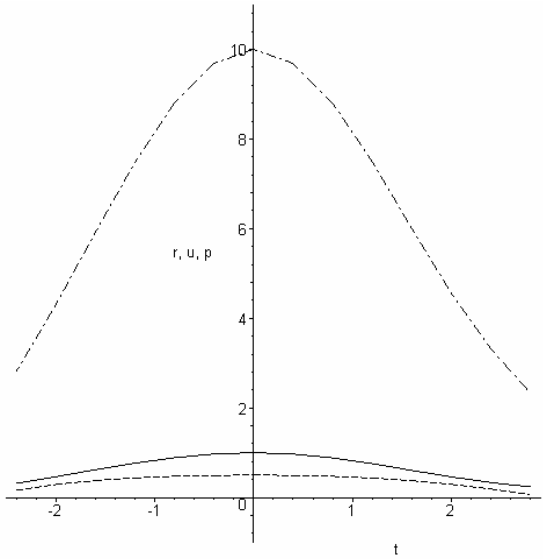

Fig. 17.

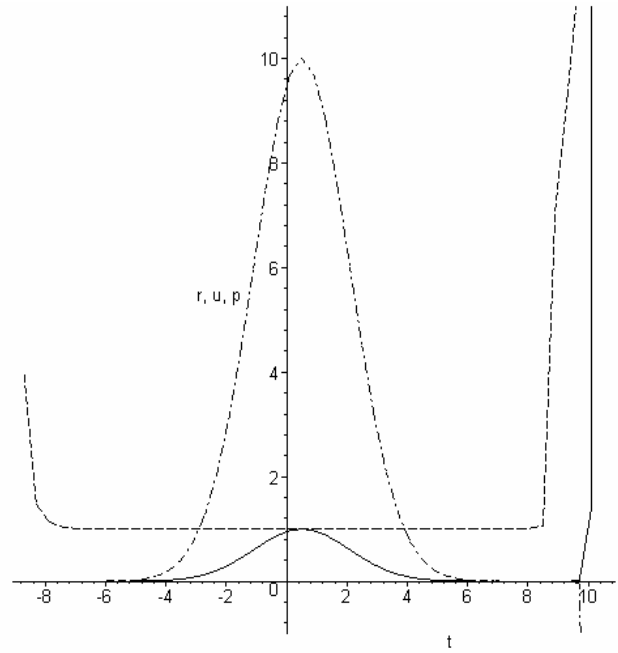

Fig. 18.

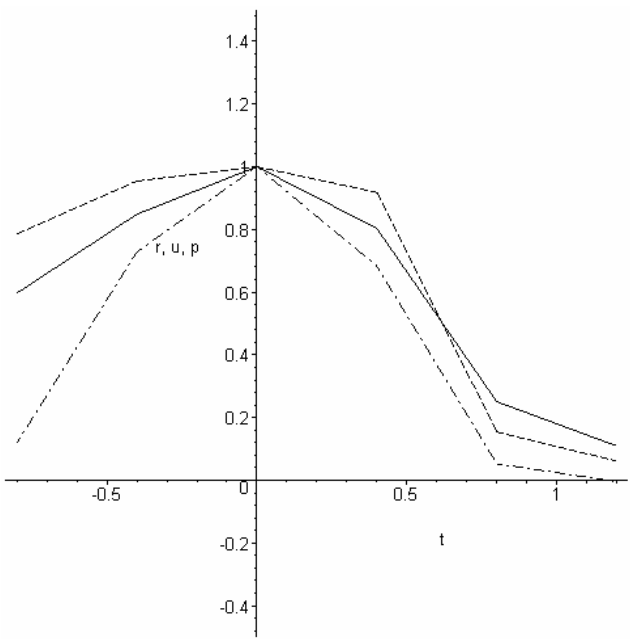

Fig. 19.

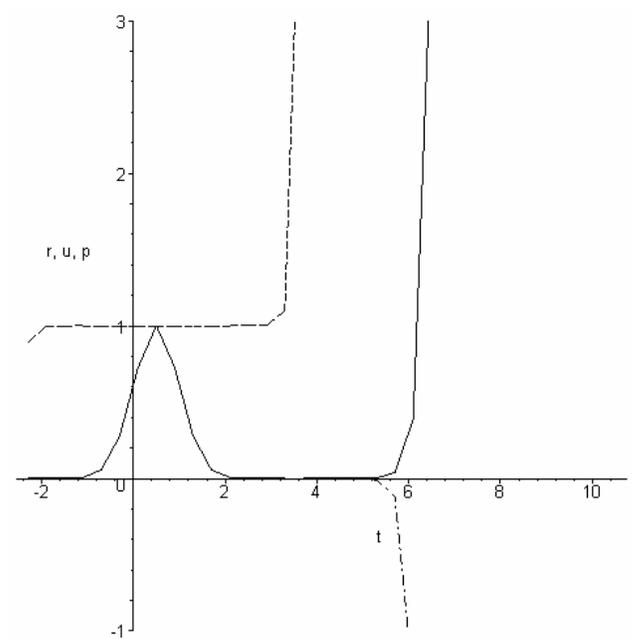

Fig. 20.

The third particular case to be considered here belongs to significant class of problems in quantum mechanics connected with the inharmonic oscillator. For such kind of oscillators potential energy $\tilde{U}$ is proportional to even degree of $\tilde{\xi}$. Figures 21 and 22 present the calculations for inharmonic oscillators: $U \cong \tilde{\xi}^{10}$ (Fig. 21) and $\tilde{U} = \tilde{\xi}^2 + \tilde{\xi}^{10}$ (Fig. 22) with the same initial conditions cited in the Maple file valid for this case of calculation

```
> dsolve[interactive]({
diff(r(t)*(1-u(t)),t)+diff((diff(p(t)+r(t)+r(t)*u(t)^2-
2*r(t)*u(t),t))/u(t)^2,t)+diff((2*t+10*t^9)*r(t)/u(t)^2,t)=0,diff(r
(t)*u(t)^2+p(t)-r(t)*u(t),t)+diff((diff(2*r(t)*u(t)^2+2*p(t)-
r(t)*u(t)-r(t)*u(t)^3-
3*p(t)*u(t),t))/u(t)^2,t)+diff((2*t+10*t^9)*r(t)/u(t)^2,t)+(2*t+10*
t^9)*(r(t)-(diff(r(t)*(u(t)-1),t))/u(t)^2)-
2*diff(r(t)*(2*t+10*t^9)/u(t),t)=0,diff(r(t)*u(t)^2+3*p(t)-
r(t)*u(t)^3-5*p(t)*u(t),t)-diff((diff(2*r(t)*u(t)^3+10*p(t)*u(t)-
r(t)*u(t)^2-3*p(t)-r(t)*u(t)^4-8*p(t)*u(t)^2-
5*p(t)^2/r(t),t))/u(t)^2,t)+diff((2*t+10*t^9)*(3*r(t)*u(t)^2+5*p(t)
)/u(t)^2,t)-2*(2*t+10*t^9)*r(t)*u(t)-
2*diff((2*t+10*t^9)*r(t)/u(t),t)+2*(2*t+10*t^9)*(r(t)*(2*t+10*t^9)+
diff(p(t)+r(t)*u(t)^2-
r(t)*u(t),t))/u(t)^2=0,u(0)=1,p(0)=1,r(0)=.1,D(u)(0)=0,D(p)(0)=0,D(
r)(0)=0});
Warning, could not obtain numerical solution at all points, plot may be
incomplete
```

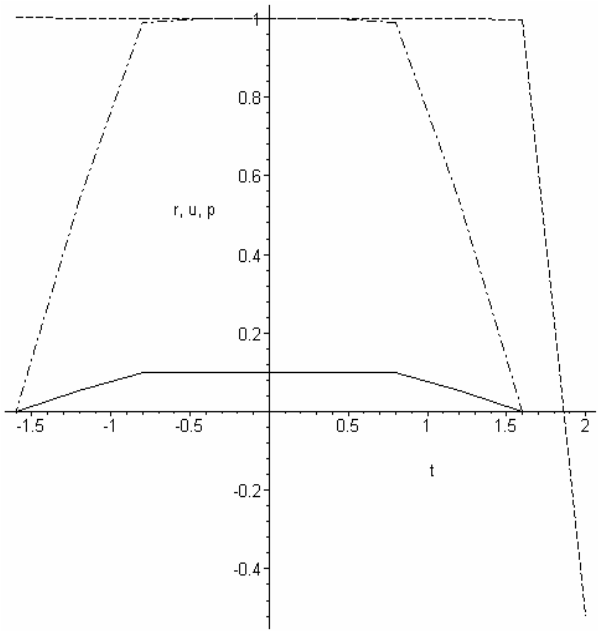

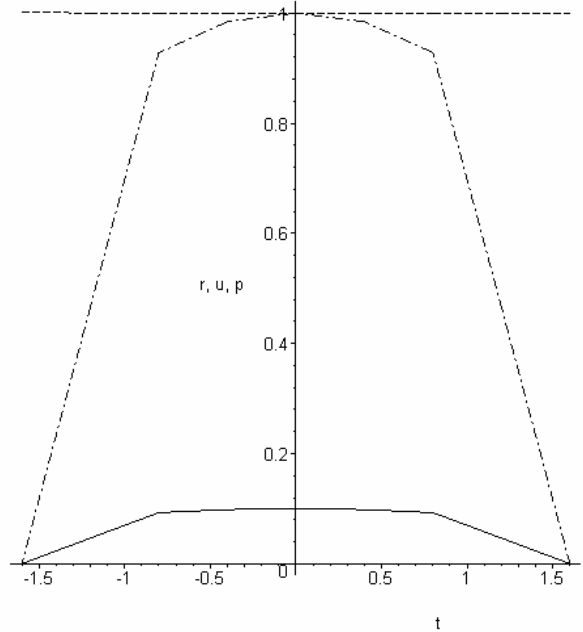

Fig. 21.        Fig. 22.

The last case has the aim to demonstrate the possibility to calculate potential energy in the process of self consistent calculation of the generalized Poisson equation

$$\frac{\partial^2 \tilde{U}}{\partial \tilde{\xi}^2} = -\left[\tilde{\rho} - \frac{1}{\tilde{u}^2}\left(-\frac{\partial \tilde{\rho}}{\partial \tilde{\xi}} + \frac{\partial}{\partial \tilde{\xi}}(\tilde{\rho}\tilde{u})\right)\right] \qquad (6.4)$$

and the system of non-local equations (6.1) – (6.3).

The following notations on Figures 23, 24 are used: $r$ - density $\tilde{\rho}$ (solid line), u- velocity $\tilde{u}$ (dashed line), $p$ - pressure $\tilde{p}$ (dash dot line), v – self consistent potential $\tilde{U}$ (dotted line). Fig. 23, 24 demonstrate the model calculations (which can be useful, for example, in the theory of Schotky contact) realized for initial perturbations, correspondingly:

```
u(0)=1,p(0)=2,r(0)=1,D(u)(0)=0,D(p)(0)=0,D(r)(0)=0,D(v)(0)=0,
v(0)=1;
u(0)=1,p(0)=1,r(0)=1,D(u)(0)=0,D(p)(0)=0,D(r)(0)=0,D(v)(0)=0,
v(0)=1.
```

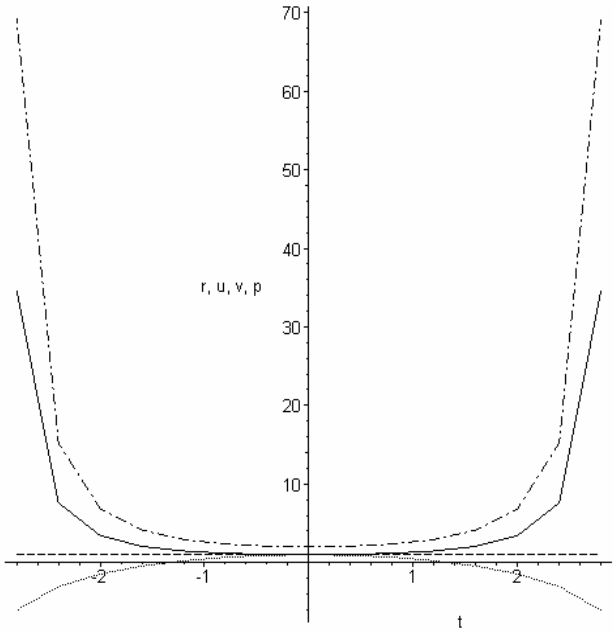
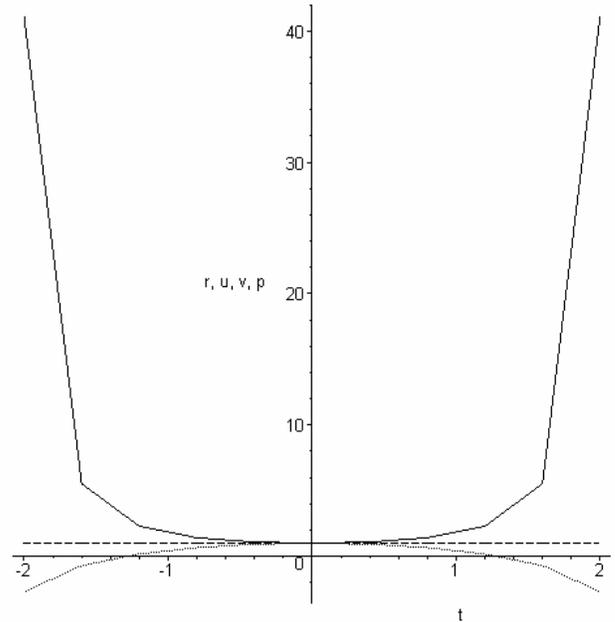

Fig. 23.    Fig. 24.

**Conclusion**

The discovery of the differential equations which can lead to the soliton's solution everywhere was the significant event in mathematics and mechanics. As we see the soliton's appearance in the generalized hydrodynamics is an "ordinary" oft-recurring fact. In many cases the theory leads to appearance of discrete levels for derivatives of hydrodynamic values. The elimination of the energy equation from consideration does not lead to the simplification of the generalized theory, moreover in many cases such kind of elimination could lead to unacceptable results from the physical point of view.

In non-local hydrodynamics solitons exist for complicated systems like translational moving inharmonic oscillator and combination of oscillators. These formations imitate the evolution of such complicated objects as molecules. The following natural step of the theory consists in investigation of colliding quantum solitons on the basement of the generalized quantum hydrodynamics. In this case we could wait for the generation of many solitons – fragments, "splinters" of process. The relativistic generalization of the delivered theory (see also [2]) allows to apply the theory for investigation of evolution of heavy particles encountering in the modern colliders.

Schroedinger quantum mechanics in many cases leads to the dissipation of the wave packets. But appearance of solitons in the generalized quantum hydrodynamics allows treating quantum effects from position of evolution of waves of matter.

Generalized hydrodynamics is very effective instrument of investigation of different physical objects in the frame of unified theory.